\documentclass[a4,10pt]{article}

\usepackage{arxiv}
\usepackage{fancyhdr}
\usepackage[utf8]{inputenc} 
\usepackage[T1]{fontenc}    
\usepackage{url}            
 \usepackage{cmbright}  

\usepackage{nicefrac}       
\usepackage{microtype}      
\usepackage{lipsum}
\usepackage{graphicx}
\graphicspath{ {./images/} }
\usepackage{tikz}   

\usepackage{cite}
\usepackage{amsmath,amssymb,amsfonts}
\usepackage{xcolor}
\definecolor{myorange}{RGB}{100, 50, 0}
\definecolor{myblub}{RGB}{34, 52, 168}
\definecolor{dg}{RGB}{64,64,64}

\usepackage[colorlinks = true,
		citecolor = blue,
		urlcolor= blue,
		linkcolor=dg,]{hyperref}
\usepackage{algorithmic}
\usepackage{textcomp}
\usepackage{xcolor}
\def\BibTeX{{\rm B\kern-.05em{\sc i\kern-.025em b}\kern-.08em
    T\kern-.1667em\lower.7ex\hbox{E}\kern-.125emX}}	

\title{Evaluating Progress in Web3 Grants: Introducing the Grant Maturity Index.}

\author{
Ben Biedermann\\
\href{mailto:bb@acurraent.com}{bb@acurraent.com}\\
\textit{Metagov Grant Innovation Lab};\\
Islands and Small States Institute \\
University of Malta\\
\href{https://orcid.org/0000-0003-1331-6517}{0000-0003-1331-6517}\\
\And
Fahima Gibrel\\
\href{mailto:mgibrel.fahima@gmail.com}{gibrel.fahima@gmail.com}\\
\textit{Metagov Grant Innovation Lab}\\
Massachusetts, United States  \\
}
 \begin{document}

 \begin{tikzpicture}[remember picture, overlay]
    \node[anchor=south east, xshift=-0.5cm, yshift=0.5cm] at (current page.south east) {
        \includegraphics[width=2cm]{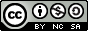} 
    };
\end{tikzpicture}

\vfill

 \pagestyle{plain}
\pagenumbering{arabic}
\maketitle

\begin{abstract}
This report introduces the Grant Maturity Index (GMI), a novel evaluative framework designed to assess the maturity and operational effectiveness of Web3 grant programs. As Web3 continues to develop, the decentralized nature of these programs brings both opportunities and challenges, particularly when it comes to governance, transparency, and community engagement. Traditional funding models are often governed by standardized processes, but Web3 grants lack such consistency, making it difficult for grant operators to measure the long-term success of their programs.

The \textit{Grant Maturity Index (GMI)} was created through \textit{exploratory applied research} to address this gap. Inspired by the World Bank’s GovTech Maturity Index (GTMI), the GMI is tailored specifically for the decentralized Web3 ecosystem. The GMI evaluates key dimensions of grant programs—governance, transparency, operational efficiency, and community engagement -- providing grant operators with a clear benchmark for assessing and improving their programs.

The primary objectives of this research are to:
\begin{itemize}
    \item Identify the structural indicators that adequately describe Web3 grant programs.
    \item Describe optimal outcomes for programs by evaluating their maturity across key operational areas.
\end{itemize}

In this report, the GMI is applied to four major Ethereum Layer 2 grant programs -- \textit{Arbitrum}, \textit{Mantle}, \textit{Taiko Labs}, and \textit{Optimism}. These case studies highlight areas where Web3 grant programs require improvement, particularly in \textit{standardizing processes}, enhancing \textit{transparency}, and increasing \textit{community participation}.

\end{abstract}



\smallskip
\noindent \textit{This work is licensed under a Creative Commons Attribution-NonCommercial-ShareAlike 4.0 International License.}

\section{Key Findings}\label{sec_1}
By applying the \textit{Grant Maturity Index (GMI)}, grant operators gain access to a structured framework that helps them assess the maturity and operational effectiveness of their Web3 grant programs. The GMI provides a baseline for evaluating governance, transparency, and community engagement, allowing operators to identify gaps and opportunities for improvement. While primarily a diagnostic tool, the GMI also offers practical insights that empower grant operators to improve program processes, ensure more consistent evaluation standards, and better align their initiatives with the evolving needs of the ecosystem.

Ultimately, this research contributes to a deeper understanding of how decentralized funding models can be optimized, helping Web3 grant programs move toward greater sustainability and impact. Nevertheless, three main findings emerged in the first analysis cycle of the GMI. These are the following.

\subsection*{Lack of standardization}\label{sec_1.1}
The absence of uniform reporting, evaluation, and governance structures hinders the ability to effectively measure program maturity and indicates low maturity across all programs. The GMI provides a structured approach to support Web3 grant programs in addressing their maturity gap by offering a toolkit for comparing grant outcomes and supporting long-term program success.

\subsection*{Transparency challenges}\label{sec_1.2}
While some programs, like \textit{Arbitrum's Short Term Incentive Program (STIP)}, offer public voting mechanisms and regular updates, there remains an overall lack of transparency in Web3 grant programs. The GMI emphasizes the need for formalized criteria for evaluating outcomes and standardized dispute resolution processes to enhance accountability.

\subsection*{Community involvement}\label{sec_1.3}
Effective community engagement is crucial to decentralized governance models, yet many Web3 grant programs struggle with communicating to applicants, grantees, and stakeholders, for example, when informing about delayed decision-making. Improving communication channels and creating more opportunities for community participation will increase both the legitimacy and operational effectiveness of these programs.

\section{Introduction}\label{sec_2}
Web3 grants represent a new approach compared to established funding practices. Most significantly Web3 grant funding often relies on decentralized governance models. Despite their potential to redefine how projects are funded, these novel funding practices are yet to be standardized and the gap towards a comprehensive evaluative framework still needs to be filled. Existing research, such as studies on decentralized science (DeSci)~\cite{1} and Web3 governance~\cite{2}, only mention selected issues, leaving a systematic approach to be desired. This lack of thorough, systematic research poses substantial challenges for Web3 grantors, DAO operators, and grantees who must often navigate these waters with limited formal guidance.

This report showcases our exploratory applied research, aiming to develop the Grant Maturity Index (GMI), a novel evaluative framework inspired by the World Bank’s GovTech Maturity Index (GTMI)~\cite{3}. The GTMI effectively assesses the integration of technology in government services with a focus on governance, citizen engagement, and operational efficiency—areas that provide a foundational model for our adaptation. The GMI integrates elements of the decentralized and dynamic elements of the Web3 ecosystem, offering grant operators a systematic tool to enhance transparency, accountability, and effectiveness in their programs. To that end, the report is guided by these two research questions:

\begin{enumerate}
    \item What indicators sufficiently describe the structure of Web3 grant programs?
    \item How can the outcomes of Web3 grant programs be optimally described?
\end{enumerate}

The report proceeds through several key sections, first examining the background and motivations behind the creation of the GMI, second detailing the development of the GMI framework adapted from the GTMI, and third applying the GMI within various Web3 grant contexts to assess its effectiveness in enhancing transparency, accountability, and operational efficiency. This exploratory approach informs both the academic and practitioners’ discourse as a resource on the specific impacts and operational metrics of grant programs, extending its utility to stakeholders within and beyond the Web3 community.

\section{Motivation and Goals}\label{sec_3}
As researchers, we understand the crucial role that Web3 grant programs play in driving innovation and growth within the crypto-economic systems and the communities running on top of them. However, a major challenge remains: the lack of standardized methods to effectively assess the effectiveness and maturity of these programs.

To address this, we have developed the Grant Maturity Index (GMI). This tool is designed to provide grant operators with a clear baseline and benchmark to measure the maturity of their programs, helping identify key areas for improvement. Our initial application of the GMI to Web3 grants aims to foster more transparent, effective, and impactful grant programs, contributing to sustained growth across the ecosystem. Our goal is to offer a toolkit that helps grant operators progressively evaluate and enhance their programs' maturity. The primary objectives for the research and the toolkit are the following:

\begin{itemize}
    \item \textit{Identifying challenges --} We pinpoint the specific obstacles that hold back the maturity of grant programs in the Web3 environment.
    \item \textit{Creating a theoretical basis --} We develop strategies that are flexible and responsive to the decentralized nature of Web3.
    \item \textit{Indicating best practices --} We provide actionable insights and best practices that can enhance both the governance and operational effectiveness of Web3 grants.
\end{itemize}

By focusing on these objectives, we aim to significantly discourse on impact measurement in Web3 grants. Through both the development and application of the GMI, we're laying the groundwork for more efficient grant management, aiming to nurture a robust and thriving Web3 ecosystem.

\section{Background}\label{sec_4}
Funding towards innovation has historically driven technological and organizational growth, providing critical support to various industries and public sectors~\cite{4,5}. However, the introduction of decentralized autonomous organizations (DAOs) to Web3 has disrupted these traditional frameworks. DAOs bring new challenges, particularly in transparency and governance, areas that are still under-researched~\cite{1,6}.

Unlike traditional grant programs, Web3 grants often operate in environments without fixed guidelines or established structures. Meanwhile, the decentralized structure and dynamic working environment of Web3 adds complexity to how funds are allocated, how risks are managed, and how technical assistance is provided. These challenges require innovative and agile management practices that reflect the unique demands of decentralized systems~\cite{7,8}.

Research in this area remains relatively underdeveloped. Although case studies and preliminary frameworks, such as quadratic funding and retroactive grants, have been explored in gray literature, these models have not been rigorously tested for their long-term effectiveness in the context of the disbursement of Web3 grants. This gap in the literature, particularly concerning the evaluation of Web3 grants’ performance and impact~\cite{9}.

The Grant Maturity Index (GMI) aims to fill this gap by taking inspiration from the World Bank’s GovTech Maturity Index (GTMI) and adapting traditional metrics for governance, fund distribution, and stakeholder engagement to the decentralized structure of Web3. The goal of the GMI is to enhance transparency, accountability, and the effectiveness of Web3 grant programs, offering a structured way to evaluate and improve these emerging funding models~\cite{3}. By providing a clear evaluative framework, the GMI addresses a critical need in the rapidly evolving Web3 space~\cite{7,10}.

The development of the GMI sets the stage for a more robust and systematic evaluation of Web3 grants. By integrating insights from both traditional and decentralized funding models, the GMI not only aims to improve the operational effectiveness of Web3 grants but also seeks to engage a broader range of stakeholders. In the long term, this framework should inform further research and development through its considerations of indicator selection and framework score calculation, fostering transparency and accountability in Web3.

\section{Methodology}\label{sec_5}
This section provides an abridged version of the approach used in our research, applied to develop and validate the Grant Maturity Index (GMI), which combines qualitative and quantitative research methods. This approach ensures that our findings are not only comprehensive but can also be  applicable and robust, addressing the real-world complexities of Web3 grant programs.

The methodology for this research was designed to develop and validate the Grant Maturity Index (GMI), combining both qualitative and quantitative research methods. This approach allows for a comprehensive analysis of Web3 grant programs, addressing their complex and evolving nature. The methodology is structured into distinct steps. The first steps involved a detailed \textit{state-of-the-art analysis} of existing maturity indices from various disciplines~\cite{11,12,13}, including sustainable supply chain management and institutional complexity in health sciences. This provided a foundation for defining the maturity-related variables relevant to Web3 grant programs.

Following the state-of-the-art analysis, the \textit{research design} was structured to assess the maturity of Web3 grant programs across several dimensions. These included governance structures, operational efficiency, community engagement, and the overall effectiveness of grants in driving the intended outcomes. This framework was applied to selected Web3 grant programs, collecting data through a combination of \textit{qualitative surveys} and \textit{quantitative web scraping} from publicly available sources.

The data collection covered grant programs from major Ethereum Layer 2 providers, including \textit{Arbitrum, Mantle, Optimism, and Taiko}, which represent the subset that was selected for evaluation. The data collection process involved reviewing documents, analyzing grant applications, and gathering additional data from platforms such as \textit{Charmverse}~\cite{14} and community forums. These data points were then used to calculate the maturity index and complemented the qualitative \textit{rubric scoring framework}.

The \textit{quantitative component} of the GMI incorporates metrics such as grant size, distribution patterns, application-to-allocation ratios, and operational transparency, while the \textit{qualitative component} involves self-assessment surveys completed by grant operators to provide insights into the governance and community engagement processes.

Finally, the research team cross-validated findings, ensuring that both qualitative insights and quantitative data aligned to provide a robust evaluation of each grant program’s maturity. This mixed-methods approach allowed for a comprehensive assessment of each program's operational capabilities, transparency, and effectiveness in achieving its goals. The results were analyzed to identify common patterns and challenges across the Web3 grant programs, providing valuable insights for future improvements.

\subsection{Development of the GMI}\label{sec_5.1}
The GMI is structured by six categories, which organize the indicators used for qualitative and quantitative scoring. The framework is designed to assess both operational and structural elements of Web3 grant programs, which allow evaluating how well grant programs meet their objectives and remain accountable to their communities.

\begin{itemize}
    \item \textit{Focus Areas and Objectives (FAO):} Examining grant sizes, types, and their integration within funding rounds.
    \item \textit{Program Structure and Organization (PSO):} Analyzing sources of funds, vesting periods, and how clearly organizational roles and processes are defined.
    \item \textit{Governance (GOV):} Assessing the clarity of program objectives, the robustness of mission statements, and the effectiveness of governance mechanisms.
    \item \textit{Effectiveness and Impact (EFI):} Evaluating criteria for success, the thoroughness of audits, and the overall impact of the programs.
    \item \textit{Transparency and Accountability (TAC):} Measuring the ratios of applications to allocations and overall operational transparency.
    \item \textit{Community Engagement (COM):} Observing the number of applicants, the duration of grants, and insights into budget allocations.
\end{itemize}

\subsection{Defining Grant Maturity}\label{sec_5.2}
In the Web3 context, \textit{maturity} is viewed as a dynamic process reflecting the evolution of grant programs through the below stages, which we hope to aid grant operators in identifying their current standing and guide future growth and improvements.

\begin{itemize}
    \item \textit{Experimental Maturity:} Where programs test new funding mechanisms and governance models.
    \item \textit{Foundational Maturity:} Basic governance and organizational structures are established.
    \item \textit{Developmental Maturity:} Programs with more structured systems and clear metrics for success.
    \item \textit{Advanced Maturity:} Highly standardized processes with robust governance and active community engagement.
\end{itemize}

\subsection{Theoretical Framework}\label{sec_5.3}
The GMI draws on established maturity models from health sciences, public sector research, and institutional complexity, adapting them to fit the unique dynamics of Web3 ecosystems. This framework incorporates both qualitative assessments and quantitative metrics to offer a complete evaluation of grant programs.

\subsection{Qualitative Assessment: Self-Assessment Survey}\label{sec_5.4}
The qualitative side of the GMI is based on a self-assessment survey that evaluates Web3 grant programs across several critical areas. This tool captures subjective insights from program operators about their governance, community engagement, and alignment with ecosystem needs. Each program is scored using a 1 (Low) to 5 (High) scale, covering areas such as clarity of objectives, governance, and transparency. These scores are aggregated into key clusters that allow for a comprehensive evaluation of program operations. Practitioners can use this rubric to evaluate their programs, identify areas for improvement, and benchmark their progress against best practices in Web3. The assessment criteria include:

\begin{itemize}
    \item \textit{Clarity of Objectives:} How clearly the program’s goals are communicated.
    \item \textit{Alignment with Ecosystem Needs:} The program’s ability to address emerging ecosystem needs.
    \item \textit{Diversity of Supported Projects:} The variety of supported projects across different verticals.
    \item \textit{Organizational Clarity:} The transparency and efficiency of the program’s structure.
    \item \textit{Governance:} The decision-making processes and overall governance structure.
    \item \textit{Community Participation and Engagement:} How effectively the community is involved in the grant process.
\end{itemize}

\subsection{Quantitative Assessment: Consolidated Metrics}\label{sec_5.5}
To complement the qualitative insights, the GMI incorporates a quantitative analysis focused on measurable metrics that reflect the operational maturity and effectiveness of the programs. Selected metrics are displayed in Table 1 below.

\begin{table}[ht!]
\centering
\begin{tabular}{l|l}
\textbf{Category} & \textbf{Description} \\ \hline
Grant Size and Distribution & This includes total, minimum, maximum, average, and median grant sizes, \\
& as well as types of grants and funding mechanisms. \\ \hline
Program Structure & Evaluating organizational clarity and the platform used for managing grant applications. \\ \hline
Evaluation and Transparency & Merging objectives, documentation, and accessibility into one category that \\
& covers how transparent programs are with their goals and evaluation criteria. \\ \hline
Operational Efficiency & Ratios such as application-to-allocation share and manager-to-applicant ratio, \\
& which show how efficiently a program operates. \\ \hline
Applicant and Grant Dynamics & Applicant numbers, grant allocation per round, and average grant duration. \\ \hline
Financial Management & The overall budget since inception, budget per round, and the ratio of operational \\
& costs to grant allocations. \\ \hline
Program Longevity and Impact & Measures how long the program has been active, total round count, and its impact \\
& on the ecosystem. \\
\end{tabular}
\label{Table:1}
\caption{Selected Metrics of the Quantitative Scoring Framework}
\end{table}

\subsection{Calculation of the GMI Composite Index}\label{sec_5.6}
The GMI composite index was developed through a structured process presented below and consists of two main components. These components are the rubric scores from the self-assessment form for each program and the final Grant Maturity Index (GMI), both presented as normalized composites. The index follows an additive approach with equal weights for all rubric indicators. Normalizing the rubric scores ensures that variations within categories, such as Community Engagement, do not disproportionately affect the final GMI score. The process for calculating the GMI involved collecting data, normalizing it using a min-max function, calculating and aggregating rubric scores, and finally normalizing them to produce the GMI composite score.

\begin{enumerate}
    \item Each metric was normalized using min-max scaling to ensure comparability across different programs and variables.
    \item Normalized scores were aggregated within their respective categories (e.g., Governance, Effectiveness) to create an overall rubric score.
    \item All rubric categories were given equal weight to provide a balanced evaluation across all factors.
    \item The rubric scores were aggregated into a final composite GMI score for each program, offering a comprehensive measure of maturity.
\end{enumerate}

\subsection{Data Collection and Validation}\label{sec_5.7}
For the selected sample consisting of documents from the Web3 grant programs of Arbitrum, Mantle, Optimism, and Taiko, the data for the indicators was collected from grant applications, supporting documentation, and lists on the Charmverse platform. Furthermore, the respective social media accounts of the awarding organizations on X, Mirror.xyz, as well as their forums were used to retrieve and aggregate primary data. The data collection and validation process can be broken down into five steps. They are described as follows.

\begin{enumerate}
    \item Gathering data from grant applications, supporting documentation, and relevant digital platforms.
    \item Confirming the accuracy of collected data through cross-referencing with additional sources.
    \item Applying the GMI scoring system to evaluate each program based on the gathered data.
    \item Conducting reviews by external experts to ensure unbiased and rigorous assessment.
    \item Compiling and finalizing the scores to produce a conclusive evaluation for each program.
\end{enumerate}

\section{Sample Selection}\label{sec_6}
The programs selected for this analysis -- Arbitrum's STIP~\cite{15}, Mantle's Grants Program~\cite{16}, Optimism’s Growth and Experiment Program~\cite{17}, and the Taiko grant program~\cite{18} -- were chosen for their shared focus on network growth, innovation, and community engagement. Their operations within the communities of Ethereum Layer 2 networks made them ideal for comparing the maturity and effectiveness of grant programs. Below are details of each program:

\subsection{Arbitrum's Short Term Incentive Program (STIP)}\label{sec_6.1}
Arbitrum's Short Term Incentive Program (STIP) distributed 71.4 million ARB to support network protocols and foster ecosystem growth. Initially capped at 50 million ARB, the program was later expanded by an additional 21.1 million ARB through the STIP Backfund to support projects that were approved but missed out on funding in the first round.

\subsection{Mantle's Grants Program}\label{sec_6.2}
Mantle's Grants Program supports projects that align with its vision of expanding and enhancing the Layer 2 ecosystem. It focuses on funding innovative solutions that drive user adoption, improve scalability, and foster decentralized application development on Mantle.

\subsection{Optimism’s Growth and Experiment Grants}\label{sec_6.3}
The Optimism Growth and Experimentation Grants, managed by the Growth Experiments Sub-Committee, focus on enhancing user engagement and driving sustainable growth on the Optimism Layer 2 network. By prioritizing experimentation and innovation, these grants align with the broader ecosystem goals of scaling adoption. As part of the larger Optimism Grants House, which is divided into the Builder and Growth and Experiment subcommittees, this program was chosen as a sample for its clear alignment of goals and operations, making it an ideal candidate for evaluating grant maturity.

\subsection{Taiko Grant Program}\label{sec_6.4}
The Taiko Grant Program is designed to drive adoption of Taiko's Layer 2 network by funding projects that enhance its infrastructure and boost ecosystem engagement. The program incentivizes innovation, supporting both emerging and established projects to expand the network’s capabilities. 

\section{Findings and Insights}\label{sec_7}
This section provides Web3 grant program operators with a practical understanding of how the Grant Maturity Index (GMI) framework can assess key operational challenges, develop theoretical frameworks for improvement, and identify best practices for enhancing governance, transparency, and community engagement. By aligning with the broader research goals, this section focuses on delivering actionable insights that practitioners can apply to optimize their grant processes, improve decision-making efficiency, and foster more effective community participation.

The Grant Maturity Index (GMI) was applied across multiple Web3 grant programs, including case studies such as Arbitrum's STIP, Mantle’s Grants Program, Optimism's Growth and Experiment Grants, and Taiko’s inaugural grant program. These programs were selected based on their relevance in fostering innovation within the Ethereum Layer 2 ecosystem. The GMI framework used both qualitative assessments (via self-assessment surveys) and quantitative data (such as grant size, evaluation timeframes, and community engagement metrics) to evaluate each program's maturity. The research was guided by the following research questions:

\begin{enumerate}
    \item What indicators effectively describe the structure of Web3 grant programs?
    \item How can the outcomes of Web3 grant programs be optimally described?
\end{enumerate}

The GMI was employed to explore how each program’s structure, governance, transparency, and community engagement contribute to its overall maturity. For example, Arbitrum STIP's decentralized, on-chain governance structure was compared with Mantle’s centralized, foundation-driven governance model, highlighting the trade-offs between operational efficiency and community involvement. These findings helped identify best practices and areas for improvement, such as the need for clearer decision-making processes and better alignment with ecosystem needs.

\section{Program Specific Findings}\label{sec_8}
The Grant Maturity Index (GMI) framework was applied across four distinct Web3 grant programs: Arbitrum, Mantle, Taiko Labs, and Optimism. These programs were evaluated across multiple rubric categories, including governance, operational efficiency, transparency, and community engagement. Below are the detailed findings from each program.

\begin{table}[h!]
\centering
\begin{tabular}{l|l|l|l|l|l}

\textbf{ID} & \textbf{Description} & \textbf{Mantle} & \textbf{Taiko} & \textbf{Optimism} & \textbf{Arbitrum STIP} \\ \hline
\textbf{GMI} & \textbf{Composite (normalized rubric scores)} & \textbf{1.1807} & \textbf{3.9312} & \textbf{3.2945} & \textbf{1.8415} \\ \hline
FAO-QN & Focus Areas and Objectives & 3.8764 & 3.8364 & 4.3498 & 5.5300 \\ \hline
PSO-QN & Program Structure and Organisation & 4.0000 & 4.5000 & 2.0000 & 2.0000 \\ \hline
GOV-QN & Governance & 2.7857 & 2.5000 & 3.5000 & 1.1429 \\ \hline
EFI-QN & Effectiveness and Impact & 3.0000 & 3.0000 & 5.0000 & 3.0000 \\ \hline
TAC-QN & Transparency and Accountability & 1.5441 & 3.0347 & 3.6180 & 1.9188 \\ \hline
COM-QL & Rubric Scoring Community Engagement & 8.0192 & 14.6587 & 11.0217 & 4.8816 \\
\end{tabular}
\caption{Selected GMI Composite Scores}
\label{Table:2}
\end{table}

\subsection{Arbitrum}\label{sec_8.1}
The Arbitrum grant program performed well in most categories, particularly in \textit{Program Structure and Organization (PSO)} and \textit{Governance (GOV)}. The decentralized nature of its governance ensures a high degree of community involvement, although this slows decision-making. In the \textit{Focus Areas and Objectives (FAO)} rubric, Arbitrum showed a balanced approach, but challenges emerged in ensuring alignment between grant allocation and ecosystem needs. In terms of \textit{Effectiveness and Impact (EFI)}, Arbitrum showed moderate success, with grant programs resulting in tangible growth and ecosystem development.

\subsection{Mantle (BitDAO)}\label{sec_8.2}
Mantle, on the other hand, displayed a more centralized structure, which contributed to high \textit{operational efficiency (PSO)} but limited community involvement. Mantle’s \textit{Focus Areas and Objectives} showed some challenges, particularly in aligning grant decisions with broader ecosystem needs, as the program leaned heavily towards consumer product grants with less diversity. Mantle scored moderately in \textit{Governance}, as the centralized decision-making allowed for quicker responses, but at the cost of transparency.

\subsection{Taiko Labs}\label{sec_8.3}
Taiko Labs struggled with diversification in its grant allocations, as most supported projects were closely tied to infrastructure development rather than community-driven initiatives. Despite this, Taiko’s \textit{governance structure (GOV)} showed promise, reflecting its ongoing efforts to engage the community and decentralize its decision-making process. However, its \textit{Effectiveness and Impact (EFI)} were rated lower due to the limited measurable success of its projects so far.

\subsection{Optimism}\label{sec_8.4}
Optimism’s \textit{Governance (GOV)} and \textit{Transparency and Accountability (TAC)} were among the highest-rated areas. The program has clearly defined structures with active community participation through the Growth Experiments Sub-Committee. Optimism’s focus on decentralization through transparent processes and grant documentation ensured a robust evaluation of project outcomes. In terms of \textit{Effectiveness and Impact (EFI)}, Optimism scored well, demonstrating the success of several funded projects, particularly those targeting user growth and network activity.

\section{Key Findings -- Focus Areas}\label{sec_9}

\subsection{Focus Areas and Objectives}\label{sec_9.1}

\subsubsection{Challenges}\label{sec_9.1.1}

Across multiple programs, there is inconsistent communication of program objectives and how they align with broader ecosystem needs. Many programs face difficulties ensuring that participants fully understand the strategic goals, often resulting in misalignment between the program's intent and the projects it funds.

\subsubsection{Findings}\label{sec_9.1.2}
\begin{itemize}
    \item Arbitrum STIP: While objectives were clear, the complexity of managing multiple rounds and back funding initiatives caused communication gaps.
    \item Mantle: Objectives were communicated but lacked specificity and depth, with a narrow focus on consumer products.
    \item Taiko: Goals were explicit but lacked transparency in how objectives were tied to decision-making processes.
\end{itemize}

\subsubsection{Recommendations}\label{sec_9.1.3}
Programs should focus on establishing and communicating clear, strategic objectives that are aligned with the current and evolving needs of the ecosystem. This ensures participants understand the program's broader intent and reduces misalignment between project selection and ecosystem goals.

\subsection{Program Structure and Organization}\label{sec_9.2}

\subsubsection{Challenges}\label{sec_9.2.1}
Operational inefficiencies, such as unclear application processes and inconsistencies in organizational frameworks, are common across Web3 grant programs. These issues impede program clarity, efficiency, and ultimately reduce the effectiveness of the program.

\subsubsection{Findings}\label{sec_9.2.2}
\begin{itemize}
    \item Arbitrum STIP: The application process was comprehensive but delayed by KYC backlogs and unclear fund distribution procedures.
    \item Mantle: Centralized structures help clarify processes, but the application process is not fully transparent, particularly around KYC onboarding.
    \item Taiko: The grant process lacked clarity, with inefficiencies in both processing and decision-making.
\end{itemize}

\subsubsection{Recommendations}\label{sec_9.2.3}
Standardizing operational procedures, improving resource management, and clearly defining roles within the organization will help reduce inefficiencies. Programs should streamline processes and make the application stages transparent and accessible to all participants.

\subsection{Governance}\label{sec_9.3}

\subsubsection{Challenges}\label{sec_9.3.1}
Decentralized governance models are designed to encourage community involvement, but they often lead to inefficiencies, communication challenges, and delays in decision-making.

\subsubsection{Findings}\label{sec_9.3.2}
\begin{itemize}
    \item Arbitrum STIP: Decentralized governance fostered community engagement but led to slow decision-making and communication breakdowns, especially during tight timelines.
    \item Mantle: The decision-making process was centralized within internal employees, limiting decentralization and introducing potential biases.
    \item Taiko: Governance was centralized, leading to vague decision-making timelines and minimal community participation.
\end{itemize}

\subsubsection{Recommendations}\label{sec_9.3.3}
Programs should implement clear governance structures that balance decentralized engagement with efficient decision-making. Formalized decision-making processes and dispute resolution mechanisms will reduce delays and foster better communication between stakeholders.

\subsection{Effectiveness and Impact}\label{sec_9.4}

\subsubsection{Challenges}\label{sec_9.4.1}
Inconsistent tracking of project outcomes and a lack of standardized evaluation criteria make it difficult to measure the long-term impact of funded projects.

\subsubsection{Findings}\label{sec_9.4.2}
\begin{itemize}
    \item Arbitrum STIP: There was limited post-funding tracking, making it hard to assess the long-term success of funded projects.
    \item Mantle: Little information is publicly available on project outcomes and success rates.
    \item Taiko: A low success rate of funded projects and limited ecosystem impact were observed, with only a small percentage of grants being funded and completed.
\end{itemize}

\subsubsection{Recommendations}\label{sec_9.4.3}
Implement structured, ongoing evaluations that measure both short- and long-term project performance. Regular post-funding assessments will enable grant programs to better understand their impact and make more informed funding decisions in the future.

\subsection{Transparency and Accountability}\label{sec_9.5}

\subsubsection{Challenges}\label{sec_9.5.1}
Many programs face challenges in maintaining transparency, particularly in their evaluation criteria and reporting processes, which undermines trust and accountability within the ecosystem.

\subsubsection{Findings}\label{sec_9.5.2}
\begin{itemize}
    \item Arbitrum STIP: Inconsistent reporting standards and informal dispute resolution methods reduced transparency, despite public updates and community voting.
    \item Mantle: Post-grant processes were opaque, with little public reporting on fund use and grantee progress.
    \item Taiko: There was a lack of public reporting on grantee progress, with updates provided only to Taiko entities.
\end{itemize}

\subsubsection{Recommendations}\label{sec_9.5.3}
Programs should develop standardized evaluation frameworks and formal accountability measures, such as regular reporting and public updates. This will enhance transparency and ensure fair assessments of funded projects across all stages of the grant lifecycle.

\subsection{Community Engagement}\label{sec_9.6}

\subsubsection{Challenges}\label{sec_9.6.1}
Communication gaps and unclear processes limit broader community involvement in governance and decision-making, reducing the diversity of input and the program's overall effectiveness.

\subsubsection{Findings}\label{sec_9.6.2}
\begin{itemize}
    \item Arbitrum STIP: While community participation was encouraged, rapid decision-making timelines and communication inefficiencies limited broader input.
    \item Mantle: There was no community participation in decision-making, and grantees were restricted from engaging with the community.
    \item Taiko: Minimal community involvement and no structured engagement mechanisms were in place.
\end{itemize}

\subsubsection{Recommendations}\label{sec_9.6.3}
Programs should expand their communication strategies and create structured feedback loops to ensure that community input is considered in decision-making. Clearer and more accessible updates will foster better engagement and ensure the community plays a more active role in governance.

\section{Common Challenges}\label{sec_10}
The evaluation of Web3 grant programs revealed several recurring challenges that indicate a low maturity and low operational efficiency. These challenges, while program-specific, reflect broader issues in the decentralized grant ecosystem. Addressing them will require both strategic adjustments and structural improvements.

One of the most significant challenges relates to \textit{decentralized governance}. While decentralization fosters community participation and shared responsibility, which in turn enhances the legitimacy of programs, it also introduces inefficiencies. For instance, Arbitrum’s decentralized governance model encourages active community involvement, but this slows decision-making and complicates communication. In such cases, formalizing decision-making processes and streamlining communication channels would significantly enhance operational efficiency while maintaining the legitimacy that decentralized governance provides.

Another key issue is \textit{transparency}, which varies widely across programs. Arbitrum’s use of public updates and community voting promotes trust within its ecosystem. However, the lack of standardized evaluation criteria and informal dispute resolution methods leaves gaps in transparency. This inconsistency was observed in other programs like Mantle and Taiko as well. Standardizing evaluation frameworks and introducing formal dispute resolution mechanisms could bridge these gaps, ensuring that grant outcomes are clearer and more transparent to all stakeholders.

\textit{Resource management} poses another significant challenge. For example, Arbitrum's milestone-based funding approach is effective in theory, but instances where disbursed grant funds are misused, such as in the case of \textit{Furucombo}, highlight the risks of fund misuse. Programs across Web3, including Mantle and Optimism, face similar issues, where the lack of oversight leads to inefficiencies or even misuse of funds. Introducing stronger oversight mechanisms, along with formalized clawback procedures, would help ensure that resources are allocated and used appropriately, enhancing the overall maturity of these programs.

\textit{Operational inefficiencies} further complicate the execution of grant programs. KYC delays, application backlog, and extended timelines were noted as common issues across programs, including Arbitrum and Taiko. These delays not only slow down the execution of grants but also affect the overall program outcomes. Streamlining the application and review processes, along with more effective initial planning, could significantly improve efficiency. Such improvements would help ensure that the programs operate in a timely and effective manner.

The presence of \textit{clear governance structures} is strongly correlated with higher program maturity. Arbitrum’s STIP Backfund demonstrates that having a transparent and well-organized governance model contributes significantly to a program’s success. Programs with well-defined governance structures, such as Optimism’s Growth Experiments Sub-Committee, see better engagement and improved decision-making processes. Defining governance structures early and maintaining clarity throughout the lifecycle of the grant program can lead to more effective and impactful outcomes.

A recurring challenge across multiple programs is the \textit{lack of standardized evaluation frameworks}. Inconsistent methods of evaluating grants lead to discrepancies in assessment and decision-making. This issue was particularly evident in Arbitrum, Mantle, and Taiko, where different metrics were used for project success, making it difficult to compare outcomes across programs. Implementing a standardized evaluation framework would improve fairness, accountability, and the overall reliability of assessments.

Lastly, \textit{effective communication} remains an ongoing challenge. Many programs struggle to keep their communities informed, leading to confusion and disengagement. Programs like Arbitrum and Mantle show that regular updates and clearer expectations are necessary to maintain community involvement. By addressing these communication gaps, programs can foster more active participation, ensuring that the community plays a more engaged role in decision-making processes.

In summary, the common challenges observed across Web3 grant programs -- governance inefficiencies, transparency issues, resource management gaps, operational hurdles, lack of standardized evaluation frameworks, and communication breakdowns -- highlight the areas where improvements are necessary. By tackling these issues, Web3 grant programs can significantly enhance their maturity, operational effectiveness, and long-term impact on the ecosystem.

\section{Limitations}\label{sec_11}
Despite the comprehensive nature of the Grant Maturity Index (GMI) framework, several limitations must be acknowledged. One significant limitation stems from the reluctance of grant managers to provide candid feedback, as many wish to avoid casting their programs in a negative light. This tendency to adhere to what is socially acceptable may result in biased self-assessments, limiting the accuracy of data collected from program operators.

Moreover, some foundations behind these grant programs are incorporated in opaque jurisdictions, which complicates efforts to obtain transparent information regarding their governance and financial management. This opacity can undermine the robustness of the GMI’s evaluations, as important aspects of program oversight remain hidden from scrutiny. Currently, the GMI does not comprehensively assess the long-term sustainability of grant programs, other than by considering the duration of their existence. There is no structured overhead assessment, such as an analysis of administrative costs, that would provide insights into the long-term viability of these programs. This gap limits the GMI's ability to forecast the sustainability of grant initiatives over extended periods.

Additionally, some grant programs function less as avenues for innovation and more as mechanisms to support ongoing operations. In certain cases, grants are utilized as procurement tools for acquiring essential products and services, rather than funding new and innovative projects. There is also evidence suggesting that grant programs can be used as vehicles for mergers and acquisitions, as seen in the communications between GMX and Arbitrum, where grants were reportedly used to support consolidation efforts. For future iterations of the GMI, extending the framework to include broader datasets, such as those suggested by Blockworks, could enhance its evaluative capacity.

To address these limitations, we are developing two anonymous surveys: one targeted at grant operators and the other at grantees and applicants. These surveys aim to gather more objective and comprehensive feedback, helping to mitigate the biases that may currently affect the GMI’s assessments.

\section{Conclusion and Learnings}\label{sec_12}
The application of the Grant Maturity Index (GMI) to multiple Web3 grant programs has given an exploratory overview of the state of grant management. A major takeaway from this analysis is the need for \textit{standardization} across all levels of program design and evaluation. Without uniform processes, it becomes difficult to consistently measure maturity. This lack of standardization hinders effective comparisons between programs and limits the ability to identify areas for improvement. Therefore, standardizing key components ensures the long-term success of Web3 grant programs. Programs like Arbitrum, Mantle, and Taiko, which show variations in transparency and evaluation metrics, highlight the need for this convergence on standardized evaluation and operating procedures.

Furthermore, \textit{transparency} remains a significant challenge across the programs evaluated. While transparency is often cited as a core value of decentralized ecosystems, many grant programs struggle to implement consistent and clear reporting procedures. Arbitrum, for example, while public updates and community voting do occur, inconsistent evaluation criteria and informal dispute resolution processes create gaps in accountability. This issue is mirrored across other programs such as Mantle and Optimism, indicating that without formalized reporting and dispute resolution mechanisms, it will be difficult to build the level of trust needed for program sustainability.

Another key learning is that \textit{community involvement} is central to decentralized governance, but many programs face barriers when it comes to communication efficiency and decision-making timelines. Programs like the one of Arbitrum, which emphasize community engagement, often experience slower decision-making processes due to their decentralized nature. Expanding and streamlining communication strategies, such as clearer channels for feedback and participation, can foster deeper community involvement, thereby improving both the legitimacy and operational effectiveness of these grant programs. Taiko and Mantle could benefit from adopting similar strategies to engage their respective communities more effectively.

The \textit{lack of post-funding evaluations} is another major gap identified in this analysis. Without structured and ongoing assessments of funded projects, it becomes difficult to measure the long-term impact and success of these grants. For instance, programs like Optimism, despite their strong governance structures, would benefit from more consistent evaluations to gauge the ecosystem-wide effects of their grants. Introducing regular post-funding evaluations into the GMI framework would not only help measure project success but also guide future funding decisions, ensuring that grant programs contribute positively to the overall health of the ecosystem. From a \textit{practice-oriented perspective}, the findings point to several actionable strategies for improving Web3 grant programs:

\begin{enumerate}
    \item \textit{Institutionalize Standardization}: Implementing consistent frameworks for reporting, evaluation, and governance will not only improve transparency but also allow for better comparison between programs, enhancing overall accountability.
    \item \textit{Enhance Transparency Mechanisms}: Formalizing processes for public updates, reporting, and dispute resolution will help close the transparency gaps currently observed in programs like Arbitrum and Mantle, fostering greater trust from participants and the community.
    \item \textit{Optimize Community Engagement}: By improving communication strategies and feedback channels, programs can ensure that the community remains actively involved in decision-making processes. This will enhance the legitimacy of decentralized governance models while improving operational efficiency.
    \item \textit{Incorporate Post-Funding Evaluations}: Introducing regular and structured post-funding evaluations will allow programs to better track the success of funded projects and make data-driven decisions for future funding rounds.
    \item \textit{Address Long-Term Sustainability}: Beyond evaluating current maturity, the GMI should be expanded to assess the long-term sustainability of grant programs. This includes analyzing overhead costs and long-term financial planning, which are areas currently under-explored in the index.
\end{enumerate}

In conclusion, the GMI offers a framework for assessing the maturity of Web3 grant programs. Incorporating post-funding evaluations and sustainability assessments will further strengthen the framework and contribute to the effective operation of Web3 grant programs.

\section*{About the Researchers}\label{sec_13}

\subsection*{Ben Biedermann}\label{sec_13.1}
Ben is a published researcher with advanced knowledge on EU cryptocurrency regulation. He is also a domain expert on digital identity and has contributed to one of the first Ethereum-based decentralized identifier (DID) methods. Through his work on digital identity he discovered his excitement for Web3 infrastructure and procurement and joined RaidGuild. After bringing several community staking solutions to production, Ben became part of the MetaGov Grant Innovation Lab and the Consortium Lead of \textit{WID3+}, researching crypto-economic procurement while reading for a PhD on small-scale digital public infrastructure at the University of Malta.

\subsection*{Fahima Gibrel}\label{sec_13.2}
Feems is a governance and grants specialist in the DAO ecosystem. She co-founded the DAO Governance Collective, serves as a steward for Hats Protocol, and manages the delegate program at Interest Protocol. She is on the Gitcoin Community Council for Community Grants and has worked with several grant programs such as Arbitrum DAO, Polygon, POKT. Feems also advises on grant frameworks and impact assessments. She holds an MSc in Lobbying and Public Affairs, with experience spanning tech, government, and global affairs.

\section*{Acknowledgment}\label{sec_14}

This work is part of the \href{https://metagov.org/}{Metagov} Grant Innovation Lab and has been funded by \href{https://ens.domains/}{ENS Domains} through its \textit{Large Grants} program.

\newpage
\pagenumbering{roman}
\appendix
\section{Case Study Comparison -- Mantle, Arbitrum}\label{sec_15}

\subsection{Introduction}\label{sec_15.1}

\begin{itemize}
    \item \textit{Mantle}: A grant program with centralized governance, managed by the Mantle Foundation, which exercises primary control over decision-making and resource allocation.
    \item \textit{Arbitrum STIP}: A decentralized grant program operating under the Arbitrum DAO, with decision-making and funding allocation subject to community votes and on-chain governance.
\end{itemize}

\subsection{Governance Structure}\label{sec_15.2}
\begin{itemize}
    \item \textit{Mantle}: Governance is centralized, with decisions made primarily by the Mantle Foundation. The community’s role in governance is minimal, and decision-makers are appointed internally.
    \begin{itemize}
        \item \textit{Maturity Level}: Early-stage governance, with limited decentralization and community input.
    \end{itemize}
    \item \textit{Arbitrum STIP}: The governance council is appointed from within the Arbitrum DAO, and funding decisions go through a transparent on-chain voting process where token holders participate.
    \begin{itemize}
        \item \textit{Maturity Level}: Advanced decentralized governance with on-chain decision-making, strong community involvement in significant decisions.
    \end{itemize}
\end{itemize}

\subsection{Community Engagement}\label{sec_15.3}
\begin{itemize}
    \item \textit{Mantle}: Community engagement is limited, with no formal community voting mechanisms in place for grant selection. The foundation has primary control over the decision-making process.
    \begin{itemize}
        \item \textit{Maturity Level}: Limited community participation, with room for growth in engaging stakeholders in decision-making.
    \end{itemize}
    \item \textit{Arbitrum STIP}: The community participates actively through voting mechanisms on grant allocation and governance decisions. Token holders have a say in treasury management and project selection.
    \begin{itemize}
        \item \textit{Maturity Level}: High community engagement with formalized voting structures, showcasing a mature and decentralized model.
    \end{itemize}
\end{itemize}

\subsection{Transparency}\label{sec_15.4}
\begin{itemize}
    \item \textit{Mantle}: Transparency is limited, with decision-making and governance largely occurring behind closed doors. Evaluation criteria and decision rationales are not publicly shared.
    \begin{itemize}
        \item \textit{Maturity Level}: Low transparency, with centralized control limiting visibility into governance and funding processes.
    \end{itemize}
    \item \textit{Arbitrum STIP}: There is strong transparency in governance, with on-chain votes and public proposals. While the community votes on large decisions, the specifics of individual project evaluations may not always be shared in detail.
    \begin{itemize}
        \item \textit{Maturity Level}: High transparency in overall governance and funding allocation, with room to improve transparency in individual project evaluations.
    \end{itemize}
\end{itemize}

\subsection{Operational Maturity}\label{sec_15.5}
\begin{itemize}
    \item \textit{Mantle}: Mantle’s centralized model allows for faster decision-making, but it lacks the community oversight and feedback mechanisms that typically support long-term program sustainability. Resource allocation is efficient, but there's less accountability.
    \begin{itemize}
        \item \textit{Maturity Level}: Intermediate maturity due to efficient operations, but lacks decentralized checks and balances.
    \end{itemize}
    \item \textit{Arbitrum STIP}: The decentralized structure provides a more resilient and transparent operational framework, but decision-making can be slower due to the need for community voting. The process ensures broader input and accountability but requires stronger community management to maintain long-term engagement.
    \begin{itemize}
        \item \textit{Maturity Level}: High maturity in terms of decentralized processes, though it faces challenges in maintaining efficiency with a larger, active community.
    \end{itemize}
\end{itemize}

\subsection{Grant Allocation and Management}\label{sec_15.6}
\begin{itemize}
    \item \textit{Mantle}: Grants are allocated internally, with the Mantle Foundation controlling decisions. This allows for rapid allocation but may lead to concerns about bias and fairness.
    \begin{itemize}
        \item \textit{Maturity Level}: Early-stage maturity, with efficient resource allocation but limited transparency and fairness concerns due to centralization.
    \end{itemize}
    \item \textit{Arbitrum STIP}: Grant allocation is conducted through a transparent process, with community votes on project proposals. The STIP process ensures accountability and broad participation but may be slower due to the need for consensus.
    \begin{itemize}
        \item \textit{Maturity Level}: High maturity in terms of transparency and fairness, though speed and efficiency are potential trade-offs.
    \end{itemize}
\end{itemize}

\subsection{Key Learnings from the Case Studies}\label{sec_15.7}
\begin{itemize}
    \item \textit{Mantle}: Centralized control allows for faster decisions but limits community involvement and transparency. This model can be efficient but may not align with the ethos of decentralized ecosystems.
    \item \textit{Arbitrum STIP}: The decentralized model fosters community trust and transparency but requires well-structured processes to maintain efficiency and engagement. Its mature governance ensures inclusivity but comes with the challenge of slower decision-making.
\end{itemize}

\subsection{Conclusion and Comparative Maturity Insights}\label{sec_15.8}
\begin{itemize}
    \item \textit{Mantle} is at an earlier stage of maturity, focusing on operational efficiency at the expense of decentralization and community engagement.
    \item \textit{Arbitrum STIP} represents a more mature, decentralized model, offering greater transparency and community involvement but facing challenges in maintaining decision-making speed and operational efficiency.
\end{itemize}

\subsection{Recommendations for Future Development}\label{sec_15.9}
\begin{itemize}
    \item \textit{Mantle}: Increase community engagement through voting mechanisms and improve transparency in governance and grant allocation to move towards a more mature, decentralized model.
    \item \textit{Arbitrum STIP}: Streamline decision-making processes without sacrificing transparency, ensuring that governance mechanisms can sustain both community participation and operational efficiency.
\end{itemize}
\newpage
\section{Indicators and Rubric Scores}\label{sec_16}
\begin{table}[ht!]
\centering
\footnotesize
\begin{tabular}{l|l|l|l}
\textbf{ID} & \textbf{Description} & \textbf{Data Type} & \textbf{Unit} \\ \hline
\textbf{FAO-QN} & \textbf{Focus Areas and Objectives} & \textbf{synthetic} & \textbf{n.a.} \\ \hline
FAO-QL & Rubric Scoring Focus Areas and Objectives & numeric & scoring \\ \hline
FAO-QN-2 & Minimum Grant Size & numeric & USD \\ \hline
FAO-QN-3 & Maximum Grant Size & numeric & USD \\ \hline
FAO-QN-6 & Evaluation Timeframe & numeric & weeks \\ \hline
FAO-QN-7 & Grant Platform & string & n.a. \\ \hline
FAO-QN-8 & Link to Grant Round(s) & string & n.a. \\ \hline
FAO-QN-9 & Grant types & numeric & scoring \\ \hline
FAO-QN-10 & Funding Type & numeric & scoring \\ \hline
\textbf{PSO-QN} & \textbf{Program Structure and Organisation} & \textbf{synthetic} & \textbf{n.a.} \\ \hline
PSO-QL & Rubric Scoring Program Structure and Organisation & numeric & scoring \\ \hline
PSO-QN-1 & Origin of Funds & numeric & scoring \\ \hline
PSO-QN-2 & Vesting Period for Fund Allocation & binary & scoring \\ \hline
PSO-QN-3 & Organizational Structure of Grantor & numeric & scoring \\ \hline
PSO-QN-4 & Grant Program Principal & numeric & scoring \\ \hline
PSO-QN-5 & Grant Program Agents & numeric & signatories \\ \hline
PSO-QN-6 & Governance Structure & string & n.a. \\ \hline
\textbf{GOV-QN} & \textbf{Governance} & \textbf{synthetic} & \textbf{n.a.} \\ \hline
QOV-QL & Rubric Scoring Governance & numeric & scoring \\ \hline
QOV-QN-1 & Grant Program Objective & numeric & scoring \\ \hline
QOV-QN-3 & Existence of Program Objective Description & numeric & scoring \\ \hline
QOV-QN-4 & Link to Program Objective & string & n.a. \\ \hline
\textbf{EFI-QN} & \textbf{Effectiveness and Impact} & \textbf{synthetic} & \textbf{n.a.} \\ \hline
EFI-QL & Rubric Scoring Effectiveness and Impact & numeric & scoring \\ \hline
EFI-QN-1 & Evaluation Criteria Public & binary & scoring \\ \hline
EFI-QN-2 & Evaluation Shared with Applicants & binary & scoring \\ \hline
EFI-QN-3 & Reference to Evaluation Criteria & string & n.a. \\ \hline
EFI-QN-4 & Grant process explained & binary & scoring \\ \hline
EFI-QN-6 & Domicile Foundation & ISO Alpha-3 & n.a. \\ \hline
EFI-QN-8 & Program Audit & numeric & scoring \\ \hline
\textbf{TAC-QN} & \textbf{Transparency and Accountability} & \textbf{synthetic} & \textbf{n.a.} \\ \hline
TAC-QL & Rubric Scoring Transparency and Accountability & numeric & scoring \\ \hline
TAC-QN-4 & Average Application to Allocation share & rational number & conversion rate \\ \hline
TAC-QN-5 & Operated by a Service Provider & binary & scoring \\ \hline
TAC-QN-6 & Program Manager to Applicant Ratio & rational number & conversion rate \\ \hline
\textbf{COM-QN} & \textbf{Community Engagement} & \textbf{synthetic} & \textbf{n.a.} \\ \hline
COM-QL & Rubric Scoring Community Engagement & numeric & scoring \\ \hline
COM-QN-1 & Minimum Applicant Count per Round & numeric & headcount \\ \hline
COM-QN-2 & Maximum Applicant Count per Round & numeric & headcount \\ \hline
COM-QN-4 & Minimum Number of Grants Allocated per Round & numeric & grant count \\ \hline
COM-QN-5 & Maximum Number of Grants Allocated per Round & numeric & grant count \\ \hline
COM-QN-7 & Minimum Grant Duration & numeric & weeks \\ \hline
COM-QN-8 & Maximum Grant Duration & numeric & weeks \\ \hline
COM-QN-11 & Time of Existence & rational number & years \\ \hline
COM-QN-12 & Round Count since Inception & numeric & rounds \\ \hline
COM-QN-13 & Number of Tracks per Round & numeric & tracks \\ \hline
COM-QN-14 & Overall Budget since Inception & rational number & USD \\ \hline
COM-QN-19 & Operations Budget per Round & rational number & USD \\ \hline
COM-QN-20 & Operations Budget to Round Budget Ratio & rational number & ratio \\ \hline
COM-QN-21 & Program Management Team Size & numeric & headcount \\ \hline
COM-QN-22 & Impact Measurement & numeric & scoring \\ \hline
COM-QN-23 & Grant Size Standardisation & binary & scoring \\ 
\end{tabular}
\caption{Indicators and Rubric Scores}
\label{Table:3}
\end{table}
\newpage
\section{Abridged GMI Quantitative Scoring Results}\label{sec_17}
\begin{table}[ht!]
\tiny
\begin{tabular}{p{3cm}|p{5cm}|p{2cm}|p{2cm}|p{2cm}|p{2cm}}
\textbf{Indicator} & \textbf{Description} & \textbf{Taiko} & \textbf{Mantle} & \textbf{Arbitrum} & \textbf{Optimism} \\ \hline
Minimum Applicant Count per Round & Measures the smallest number of applicants in a round, indicating the base level of program outreach. Lower numbers suggest limited awareness or accessibility. & 118,300 & 16,000 & 98 & 15 \\ \hline
Maximum Applicant Count per Round & Tracks the highest number of applicants, reflecting program popularity and reach. High numbers suggest broad appeal and effective outreach. & 232,600 & 27,000 & 124 & 46 \\ \hline
Average Applicant Count per Round & Provides a balanced view of the program’s applicant base across multiple rounds, helping assess consistency in engagement. & n.a. & n.a. & 111 & 30.5 \\ \hline
Minimum Number of Grants Allocated per Round & Indicates the smallest number of grants given per round, showing how selective the program is in awarding funding. & 9 & 10 & 26 & n.a. \\ \hline
Maximum Number of Grants Allocated per Round & Shows the upper limit of funding allocations, indicating how much the program can or chooses to distribute in a single round. & 238 & 35 & 29 & n.a. \\ \hline
Average Number of Grants Allocated per Round & Offers insight into the typical number of grants distributed, reflecting the scale of the program. & n.a. & n.a. & 42 & n.a. \\ \hline
Minimum Grant Duration (in months) & Measures the shortest grant period, indicating the speed at which funding is disbursed and utilized. & 4 & 4 & 3 & 7.9 weeks \\ \hline
Maximum Grant Duration (in months) & Captures the longest duration of grants, showing how long projects can take to complete. & 6 & 6 & 4 & 18 weeks \\ \hline
Average Grant Duration (in months) & Reflects the typical duration of funded projects, providing insights into project timelines. & 6 & 6 & 4 & 18 weeks \\ \hline
Existence in years & Indicates how long the grant program has been running, providing context on maturity and experience. & 1 year & n.a. & <1 year & 1.5 years \\ \hline
Round Count since Inception & Measures the number of rounds completed since the program started, offering a sense of activity level. & 15 & 2 & 2 (STIP \& Backfund) & n.a. \\ \hline
Overall Budget since Inception & The total amount of funding allocated since the program's launch, reflecting its financial scale. & \$276m & \$100m & 71.4M ARB & 6.5M \\ \hline
Minimum Grant Size & Shows the smallest grant amount provided, indicating the baseline support offered to projects. & \$5,000 & \$5,000 & 38,000 ARB & <50K OP \\ \hline
Maximum Grant Size & Reflects the largest grant provided, showcasing the program's ability to fund bigger projects. & \$276m & \$25,000 & 12m ARB & 250K OP \\ \hline
Average Grant Size & Provides the average amount of funding per grant, helping to assess typical project funding. & n.a. & \$823,077 ARB & 823,077 ARB & 99,877 OP \\ \hline
Evaluation Timeframe (in weeks) & Time allocated for evaluating applications, offering insight into the program's efficiency. & 4 & 4 & 2 & 4 \\ \hline
Organizational Structure of Grantor & Describes the governing body responsible for administering the grants (e.g., DAO, Foundation). & DAO + Foundation & British Virgin Islands (Foundation) & DAO + Foundation & OP Foundation \\ \hline
Governance Structure & Evaluates the decision-making process, focusing on who allocates funds and approves grants. & 1 (Principal allocates) & 1 (Principal allocates) & 1 (Principal allocates) & 1 (Principal allocates) \\ \hline
Grant Size Standardisation & Determines if grants are standardized in size, ensuring uniformity in funding allocations. & 0 & 1 & 0 & 1 \\ \hline
Grant Types & Identifies the type of grants awarded, helping assess flexibility and risk management. & 2 (milestones) & 2 (milestones) & 2 (milestones) & 2 (milestones) \\ \hline
Funding Type (simplified) & Specifies the currency or asset used in funding, reflecting the program's financial strategy (e.g., Native Token, Stablecoin). & Native Token & Native Token & Native Token & Native Token \\ \hline
Market capitalisation of funding asset at round start & Tracks the market value of the funding asset (if using tokens), showing the financial capacity. & \$3.2B & \$121M & \$3.2B & \$2.6B \\ \hline
Evaluation Criteria Public (0 No, 1 Yes) & Measures whether evaluation criteria are made public, promoting transparency. & 1 & 1 & 1 & 1 \\ \hline
Evaluation Shared with Applicants (0 No, 1 Yes) & Indicates whether applicants are informed of the criteria, enhancing trust in the process. & 0 & 0 & 0 & 1 \\ \hline
Grant process explained & Shows if the grant application process is clearly explained, ensuring accessibility and fairness. & 1 & 1 & 1 & 1 \\ \hline
Average Application to Allocation share & Compares the number of applications received to the number of grants awarded, indicating program competitiveness. & 0.000721 & 0.0016875 & 1.644 & tbc \\ \hline
Program Manager to Applicant Ratio & Assesses the workload of program managers by comparing the number of applicants per manager. & n.a. & n.a. & 1:28 & n.a. \\ \hline
Operations Budget per Round & The amount of funding set aside for operations per round, showing the efficiency of fund management. & \$156,000,000 & \$21,000,000 & 114,000 ARB & 440k OP \\ \hline
Impact Measurement & Tracks how project success is measured, indicating the program's focus on outcomes. & 1 & 1 & 1 & 2 \\ \hline
Origin of Funds & Identifies the source of funding, showing the program's financial foundation. & 1 (DAO Treasury) & 1 (DAO Treasury) & 1 (DAO Treasury) & 0 (Token Launch) \\ \hline
Vesting Period for Fund Allocation & Determines if funds are released over time, showing how risk and reward are managed. & No & No & No & No \\ \hline
Organizational Structure of Grantor & Describes how the grantor is structured, indicating levels of oversight and accountability. & DAO + Foundation & DAO + Foundation & DAO + Foundation & Optimism Foundation \\ \hline
Funding Type & Clarifies the form of funding used, helping assess stability and flexibility. & 1 (Native Token) & 1 (Native Token) & 1 (Native Token) & 1 (Native Token) \\ \hline
Market capitalisation of funding asset at round start & Tracks the market value of the funding asset (if using tokens), showing financial capacity. & \$3.2B & \$121M & \$3.2B & \$2.6B \\ \hline
Grant Program Objective & Identifies the primary focus of the program, whether for token growth, profit, or philanthropy. & 1 (Token and network growth) & 1 (Token and network growth) & 1 (Token and network growth) & 1 (Token and network growth) \\ \hline
Link to Program Objective & Provides a link to the program’s objective, ensuring transparency and accountability. & Link & Link & Link & Link \\ \hline
Domicile Foundation & Indicates the jurisdiction of the foundation, showing the legal framework for operations. & Cayman Islands & British Virgin Islands & Cayman Islands & Cayman Islands Foundation \\ \hline
Grant Program Principal & Identifies the main entity responsible for running the program. & DAO & DAO & DAO & Optimism Foundation \\ \hline
Grant Program Agent & Lists the agents who manage the day-to-day operations of the grant program. & 7 (tnorm and 6 signers) & 7 (tnorm and 6 signers) & 7 (tnorm and 6 signers) & 2 \\ \hline
Governance Structure & Evaluates how the program is governed, showing the decision-making process. & 1 (Principal allocates) & 1 (Principal allocates) & 1 (Principal allocates) & 1 (Principal allocates) \\ 
\end{tabular}
\caption{Abridged GMI Quantitative Scoring Results}
\label{Table:4}
\end{table}


\begin{thebibliography}{00}
\bibitem{1} W. Ding \textit{et al.}, ``DeSci Based on Web3 and DAO: A Comprehensive Overview and Reference Model'', IEEE Trans. Comput. Soc. Syst., vol. 9, no. 5, pp. 1563--1573, Oct. 2022, doi: \href{https://doi.org/10.1109/TCSS.2022.3204745}{10.1109/TCSS.2022.3204745}.
\bibitem{2} D. W. E. Allen, C. Berg, A. M. Lane, T. MacDonald, and J. Potts, ``The Exchange Theory of Web3 Governance'', Sep. 05, 2022, \textit{Rochester, NY}: 4209827. doi: \href{https://doi.org/10.2139/ssrn.4209827}{10.2139/ssrn.4209827}.
\bibitem{3} C. Dener, H. Nii-Aponsah, L. E. Ghunney, and K. D. Johns, ``GovTech Maturity Index: The State of Public Sector Digital Transformation'', World Bank Publications, 2021. \href{https://doi.org/10.1596/978-1-4648-1753-7}{10.1596/978-1-4648-1753-7}.
\bibitem{4} J. Albors and R. Barrera, ``Impact of Public Funding on a Firm's Innovation Performance: Analysis of Internal and External Moderating Factors'', Int. J. Innov. Manag., vol. 15, pp. 1297--1322, Dec. 2011, doi: \href{https://doi.org/10.1142/S136391961100374X}{10.1142/S136391961100374X}.
\bibitem{5} J. R. Bartle and R. L. Korosec, ``A Review of State Procurement and Contracting'', J. Public Procure., vol. 3, no. 2, pp. 192--214, Jan. 2003, doi: \href{https://doi.org/10.1108/JOPP-03-02-2003-B003}{10.1108/JOPP-03-02-2003-B003}.
\bibitem{6} F. Monteiro and M. Correia, ``Decentralised Autonomous Organisations for Public Procurement'', in Proceedings of the 27th International Conference on Evaluation and Assessment in Software Engineering, in EASE '23. New York, NY, USA: Association for Computing Machinery, Jun. 2023, pp. 378--385, doi: \href{https://doi.org/10.1145/3593434.3593519}{10.1145/3593434.3593519}.
\bibitem{7} V. Shermin, ``Disrupting Governance with Blockchains and Smart Contracts'', Strateg. Change, vol. 26, no. 5, pp. 499--509, Sep. 2017, doi: \href{https://doi.org/10.1002/jsc.2146}{10.1002/jsc.2146}.
\bibitem{8} K. Gilbert, B. Tenni, and G. Lê, ``Sustainable Transition From Donor Grant Financing: What Could It Look Like?'', Asia Pac. J. Public Health, vol. 31, no. 6, pp. 485--498, Sep. 2019, doi: \href{https://doi.org/10.1177/1010539519870656}{10.1177/1010539519870656}.
\bibitem{9} [1] E. Leventhal, M. Waqar, A. Liu, B. Biedermann, F. Gibrel, H. Devjani, L. NaMu, M. Grendel, and V. Elefante, ``State of Web3 Grants Report 2024''. 2024. doi: \href{http://dx.doi.org/10.13140/RG.2.2.19305.51048}{10.13140/RG.2.2.19305.51048}.
\bibitem{10} S. T. Howell, ``Financing Innovation: Evidence from R\&D Grants'', Am. Econ. Rev., vol. 107, no. 4, pp. 1136--1164, Apr. 2017, doi: \href{https://doi.org/10.1257/aer.20150808}{10.1257/aer.20150808}.
\bibitem{11} E. S. Barry, J. Merkebu, and L. Varpio, ``State-of-the-art Literature Review Methodology: A Six-Step Approach for Knowledge Synthesis'', vol. 11, no. 5, Art. no. 5, Sep. 2022, doi: \href{https://doi.org/10.1007/S40037-022-00725-9}{10.1007/S40037-022-00725-9}.
\bibitem{12} A. Kucińska-Landwójtowicz, I. D. Czabak-Górska, P. Domingues, P. Sampaio, and C. Ferradaz de Carvalho, ``Organizational Maturity Models: The Leading Research Fields and Opportunities for Further Studies'', Int. J. Qual. Reliab. Manag., vol. 41, no. 1, pp. 60--83, Jan. 2023, doi: \href{https://doi.org/10.1108/IJQRM-12-2022-0360}{10.1108/IJQRM-12-2022-0360}.
\bibitem{13} E. Yatskovskaya, J. S. Srai, and M. Kumar, ``Integrated Supply Network Maturity Model: Water Scarcity Perspective'', Sustainability, vol. 10, no. 3, Art. no. 3, Mar. 2018, doi: \href{https://doi.org/10.3390/su10030896}{10.3390/su10030896}.
\bibitem{14} NorthShore.ai Inc., ``Web3 Grants. Highly Configurable, End-to-end Management Solution'', CharmVerse | The Network for Onchain Communities. Accessed: Aug. 10, 2024. [Online]. Available: \href{https://charmverse.io/solutions/grants/}{https://charmverse.io/solutions/grants/}.
\bibitem{15} Arbitrum DAO, ``Arbitrum Short-term Incentive Program'', Arbitrum DAO Proposal. Accessed: Aug. 10, 2024. [Online]. Available: \href{https://snapshot.org/\#/arbitrumfoundation.eth/proposal/0x5e0057920df9a278918e4de4ee1e6de7e2415d2af985e40a6c7a1a4b47a4ce01}{https://snapshot.org/\#/arbitrumfoundation.eth/proposal/0x5e0057920df9a278918e4de4ee1-e6de7e2415d2af985e40a6c7a1a4b47a4ce01}.
\bibitem{16} BitDAO, ``[PASSED] MIP-24: Mantle EcoFund'', Mantle Forum. Accessed: Aug. 10, 2024. [Online]. Available: \href{https://forum.mantle.xyz/t/passed-mip-24-mantle-ecofund/4692}{https://forum.mantle.xyz/t/passed-mip-24-mantle-ecofund/4692}.
\bibitem{17} The Optimism Collective, ``What is the Optimism Collective? Get a Grant'', Optimism Docs. Accessed: Aug. 09, 2024. [Online]. Available: \href{https://community.optimism.io/docs/governance/get-a-grant.html}{https://community.optimism.io/docs/governance/get-a-grant.html}.
\bibitem{18} Taiko Labs, ``Announcing our First Community Grant Program''. Accessed: Aug. 10, 2024. [Online]. Available: \href{https://taiko.mirror.xyz/G7dmuoR42S4D55vT8bs_lAxPZP63kAgRu2IfqkJdf6U}{https://taiko.mirror.xyz/G7dmuoR42S4D55vT8bs\_lAxPZP63kAgRu2IfqkJdf6U}.

\end{thebibliography}
\end{document}